\documentclass[preprint,times]{emulateapj}
\usepackage{graphicx}
\usepackage{url}
\usepackage{natbib}
\usepackage{epsf}
\usepackage{color} 

\begin{document}

\title{Extreme AO Observations of Two Triple Asteroid Systems with SPHERE}
\author{B. Yang$^{1,2}$, Z. Wahhaj$^{1}$, L. Beauvalet$^{3}$, F. Marchis$^{4}$, C. Dumas$^{1,5}$, 
          \and M. Marsset$^{1}$, E. L. Nielsen$^{4,6}$, F. Vachier$^{7}$}
\affil{$1$ European Southern Observatory, Chile\\
 $2$ Yunnan Astronomical Observatory, Chinese Academy of Sciences, China\\
 $3$ National Observatory, Rio de Janeiro, Brazil\\
 $4$ Carl Sagan Center at the SETI Institute, USA\\
 $5$ TMT International Observatory, USA\\
 $6$ Kavli Institute for Particle Astrophysics and Cosmology, Stanford University, Stanford, CA 94305, USA\\
 $7$  Institut de M\'ecanique C\'eleste et de Calcul des \'Eph\'em\'erides, France}
\email{byang@eso.org}

 
\begin{abstract}
We present the discovery of a new satellite of asteroid (130) Elektra - S/2014 (130) 1 - in differential imaging and in integral field spectroscopy data over multiple epochs obtained with SPHERE/VLT. This new (second) moonlet of Elektra is about 2 km across, on an eccentric orbit and about 500 km away from the primary. For a comparative study, we also observed another triple asteroid system (93) Minerva. For both systems, component-resolved reflectance spectra of the satellites and primary were obtained simultaneously. No significant spectral difference was observed between the satellites and the primary for either triple system. We find that the moonlets in both systems are more likely to have been created by sub-disruptive impacts as opposed to having been captured.
\end{abstract}

%

\section{Introduction}
Asteroids are the relics of the building blocks that formed the terrestrial planets in the early days of the Solar System. Asteroids with satellites are of particular importance because their formation mechanisms, accretional and collisional processes, are critical in planet formation and evolution. In addition, multiple asteroids provide otherwise unattainable information about the intrinsic properties of the system: composition, interior structure and evolutionary processes \citep{margot:2015}. Recently, a second-generation extreme-AO instrument, Spectro-Polarimetric High-contrast Exoplanet Research instrument  \citep[SPHERE;][]{beuzit:2008}, has been commissioned at one of European Southern Observatory's  8~m Very Large Telescopes (VLT) at Cerro Paranal in Chile. SPHERE, with its unprecedented sensitivity and spatial resolution, offers a fresh opportunity for detecting and characterizing multiple asteroid systems.

More than a hundred binary or triple asteroids have been detected. However, how and under what conditions these systems formed is poorly understood. Component-resolved spectroscopy of the primary and the moonlet obtained simultaneously is the key to understanding their formation. If the moonlets resulted from an impact, then they would be of the same composition as the primary. Whereas, if the multiple system originated from capture during a three-body encounter, then the composition of the satellites would differ from that of the primary. Most of the available component-resolved spectroscopic studies find no significant differences between the primary and secondary, for example (22) Kalliope \citep{laver:2009}, (90) Antiope \citep{marchis:2011}, (379) Huenna \citep{deMeo:2011} and (809) Lundia \citep{birlan:2014}. However, a recent work on two triple asteroid systems, (45) Eugenia and (87) Sylvia, yield rather surprising results. Using SINFONI/VLT, \cite{marchis:2013a} found that the spectra of the moonlets are noticeably redder than their primaries. The difference in the spectral slopes of the individual components for Eugenia and Sylvia is not well understood. One possible explanation is that the primary and the satellites may have experienced different surface alteration processes (such as space weathering effects; see Lantz et al.\ 2013\nocite{lantz:2013}) due to the difference in their dynamical ages. Alternatively, satellites may not result from impact but originate, instead, from a capture during a three-body encounter \citep{funato:2004}. To investigate these possibilities, we performed direct imaging and integral field spectroscopy of two multiple asteroids (i.e.\ (93) Minerva and (130) Elektra) using SPHERE.
 
(93) Minerva is the fifth triple asteroid discovered in the Main Belt \citep{marchis:2009}. The primary, located in the mid asteroid belt ($r$=2.75 AU), is a large and dark asteroid (D = 147$\pm$2 km, p$_v$ = 0.068 $\pm$ 0.003, Usui et al.\ 2011\nocite{usui:2011}) and its spectrum is classified as a C-type \citep{lazzaro:2004, deMeo:2009}. (130) Elektra was one of the four binary systems with eccentric mutual orbits \citep{marchis:2008b}. The primary is also a large and dark main belt asteroid (D = 197$\pm$20 km, p$_v$ = 0.064 $\pm$ 0.013, Marchis et al.\ 2012\nocite{marchis:2012}). Its optical spectrum is classified as a G or Ch-type \citep{tholen:1989, bus:2002}. Both Minerva and Elektra are low-albedo primitive asteroids, as are Eugenia and Sylvia. If the space weathering effect is responsible for the spectral difference between the primary and the two satellites seen in the Eugenia and Sylvia systems, the same process may have also modified the surface properties of the Minerva and Elektra systems. If so, we would expect the spectra of the moonlets to be different from those of the primaries.


\section{Observation and Data Reduction}
Observations were carried out with the two infrared subsystems of SPHERE/VLT simultaneously, i.e.\ the infrared differential imager and spectrograph (IRDIS; Dohlen et al.\ 2008) and the integral field spectrograph \citep[IFS;][]{claudi:2008}.
Low resolution spectra were obtained with the IFS from 0.95-1.65 $\mu$m (Y to H-band,  spectral resolution $\sim$30), while simultaneous images were obtained in two narrow  ($\Delta \lambda$ = 0.1~$\mu$m) bands K1 and K2 ($\lambda$=2.11~$\mu$m and 2.25~$\mu$m) with IRDIS. For (130) Elektra, we observed in the field stabilized mode, where the sky remains fixed with respect to the detector.  (93) Minerva was observed in the pupil-tracking mode \citep{liu:2004,marois:2006}, where the sky rotates, but the pupil stays fixed with respect to the detector. The fields of view (FOV) of IRDIS and IFS are 11$\arcsec \times$12.5$\arcsec$ and 1.73$\arcsec \times$1.73$\arcsec$ respectively, while the pixel scale is 0.0123$\arcsec$ for both. 

We used the SPHERE consortium's pipeline \citep{pavlov:2008} for data reduction. 
See \cite{mesa:2015} and \cite{vigan:2015} for further details.
IRDIS images were dark subtracted, flat fielded and bad pixels removed.
The two K-band channels were combined to improve the signal-to-noise ratio (SNR). 
The IFS data were dark subtracted, bad pixel treated, flat fielded and wavelength calibrated. 
The data were then resampled into a cube of 39 images of 3.3\% band width ($\Delta \lambda / \lambda$) over 
the spectral range and with a scale of 0.0074$\arcsec$ per spaxel. Isolated bright and dark pixels were found in the IFS data which is due to imperfections in the wavelength calibration procedure. The IDL routine {\it acre} (by Marc Buie) was used to remove these via interpolation between neighboring pixels. 

Both systems were observed between UT December 6 and 9, 2014 as part of SPHERE science verification.
Additional observations on Elektra were obtained on UT December 30 and 31, 2014, using director's discretionary time. 
At each epoch, we also observed a nearby solar analog (G2V) star to enable effective removal of telluric absorption artifacts and the solar color gradient. This star also served as a point-spread-function (PSF) estimate for each spectral channel.  

\section{Additional Image Processing}
The known moonlet of (130) Elektra is not far from the primary and is nearly buried in the primary's halo. If this halo is not removed carefully, it could contaminate the photometry of its moonlet. Below, we describe the technique used to minimize the contamination from the primary. A two-dimensional Gaussian was fitted to the primary to determine its optical center. The primary's halo was then removed beyond $\sim$0.15$''$ from the primary using a process described in \cite{wahhaj:2013}. In this method, local medians over arcs (centered on the primary) of length 15--40 pixels are removed from each pixel. The arc lengths used depended on spatial scale of background structure, but were at least three times the full-width-half-maximum (FWHM) of any field source. 

One unavoidable side effect of this process is that the flux of a moonlet is partially subtracted along with the halo. To estimate this signal loss, we performed simulations on the science data set. We inserted Gaussian point sources into the reduced IFS images for wavelength. The surface of the moonlets of neither system were resolved by SPHERE and their PSFs, which depended on the AO performance, had FWHMs close to the diffraction limit. The FWHM of the simulated sources at each spectral channel was determined by PSF fitting of the  standard star that was observed right before or after the asteroid observations. We then defined annuli centered on the primary, extending 10 pixels radially inwards and outwards from each satellite. In each annulus, we randomly selected 200 positions and inserted two types of point sources at each position. A strong source with 10 times the estimated flux of the satellite was used to estimate the systematic flux loss.  A weaker source with a flux similar to that of the satellite was used to estimate the random error of our photometric measurements. After removing the primary's halo, we measured the brightness of the recovered sources using  two methods: aperture photometry and PSF fitting. Our analysis showed that the PSF fitting method generally yields smaller photometric uncertainties. However, when the moonlets were too faint or exhibit an irregular shape due to high airmass or poor AO correction, the PSF fitting method failed to emulate the PSF shape. In these cases, aperture photometry was used instead. 

\section{Discovery of New Satellite S/2014 (130) 1}  
As a result of this process (see Figure 1), a new moonlet, S/2014 (130) 1, was detected both in the IFS and IRDIS data sets over several epochs. The strongest signal of the new moonlet was detected on UT December 9 when it was 0.36$^{\prime\prime}$ from the primary. On December 31, the new inner satellite was only detected in the IFS data. The weakened signal of the inner satellite was due to the less favorable geometry where the asteroid system was farther away from the Sun and the Earth. The relative positions of the two satellites from the primary are listed in Table 1. The diameters of the satellites, estimated by calculating the integrated flux ratio of the primary to each companion, are also listed in Table 1. The effective diameter of the primary of 197 $\pm$ 20 km is adopted from \cite{marchis:2012}. Assuming that the satellites have the same albedo as the primary, the estimated diameter of the new moon, S/2014 (130) 1, is 2.0$\pm$1.5 km and the estimated diameter for S/2003 (130) 1 is 6.0$\pm$1.5 km. 

\subsection{Numerical modeling and fitting}
We obtained the orbital parameters of the two satellites of (130) Elektra using the numerical code ODIN \citep{beauvalet:2013}, which simultaneously integrates the heliocentric motion of the system, and the motion of the individual components around their barycenter.  ODIN has been used to study the Pluto system and the triple systems, such as (45) Eugenia and (87) Sylvia \citep{beauvalet:2014}. Using ODIN to fit Elektra's astrometric data and assuming that the masses of the satellites are negligible, we obtained satisfactory orbital solutions for the two moonlets, see Figure 2a. For satellite S/2003(130) 1, we also included the previous observations from \citet{marchis:2008b} and compared our results with those obtained by GENOID \citep{vachier:2012}. We found that the two models provided very similar results. The orbital periods for the inner and the outer moonlets are 1.256$\pm$0.003d and 5.287$\pm$0.001d, respectively. The residuals for the outer moonlet in $\alpha*\cos(\delta)$ and $\delta$ are: 20 mas and 25 mas, where $\alpha$ is right ascension and $\delta$ is declination. For the inner satellite, S/2014 (130) 1, the residuals in RA and DEC are 11 mas and 14 mas, respectively. We used the random hold-out method described in \cite{marchis:2010} to estimate the uncertainties of the orbital elements. We randomly removed 3 observations from the dataset and fit the remaining observations to obtain an orbital solution. We repeated this process 100 times and adopted 3-$\sigma$ uncertainty for each orbital element. We attempted to fit the gravitational harmonic parameter $J_2$, which is the main indicator of the mass distribution of the primary. We adopted an initial value of 0.13, based on axial ratios from shape model estimates \citep{marchis:2008b} and assumed a homogeneous interior. However, we found reasonable orbital solutions for a wide range of J$_2$ values from 0 to 0.13. More astrometric observations, especially of the inner moonlet, are needed to constrain the J$_2$ of the Elektra system.

\subsection{Stability}
 We checked the stability of the orbits of the both satellites of Elektra using ODIN, the accuracy of which has been tested using other well-observed systems by simulating them for $>$100 years. Our main goal is to verify whether the orbits of the two moonlets are stable between 2006 and 2014. We performed two different simulations over 20 years. Neither of these simulations included the Yarkovsky, YORP or tidal effects, which are only needed for very-long term integrations. In one simulation, the masses of the satellites are negligible. In the other simulation, the two satellites have masses of $7 \times 10^{12}$ kg and $2\times 10^{14}$ kg respectively,  assuming the same density as the primary of 1.7 g/cm$^3$ \citep{marchis:2008b}.  Neither simulation showed significant instability over time. The semi-major axis for S/2014 (130) 1 has a small oscillation spanning less than 100 meters. For  S/2003 (130) 1, semi-major axis oscillates less than 1.4 km. 
 
\section{Component-Resolved Spectroscopy}  
\subsection{The (130) Elektra system}
 The Elektra system was observed on four nights.  However, we have reliable spectra for the inner moonlet from December 09 and December 30 only. On UT December 9, the primaryÕs halo varied significantly in intensity over small spatial scales near the inner moonlet. Hence we used an arc length of 15 pixels for halo-removal. The solar analog star, HD20926, had an average PSF FWHM over 39 spectral channels of 5.3 pixels. The outer moonlet is roughly 0.6$^{\prime\prime}$ away from the primary where the primary's halo is more uniform and so an arc of length 30 pixels was used to remove the background flux. On UT December 30, given a different AO performance, an arc of length 40 pixels was used for the both moonlets for the halo removal. The solar analog star, HD19315, had an average PSF FWHM of 4.9 pixels. The spectra of the Elektra system are presented in Figure 3, which have been corrected for the systematic flux loss estimated from our simulations. We calculated the spectral slope, s$'$, the fractional intensity change per 1000 \AA, using a weighted linear regression. We performed students' t-test and calculated P to compare the spectra of the satellites and that of the primary, where P is used as an indicator to determine whether the spectra of the individual component are significantly different. We found P=0.56 and 0.82 on UT December 09 and 30 respectively for the inner moonlet and P=1.0 and 0.56 for the outer moonlet. Since the p-values are much larger than the significance level of 0.05, we conclude that there is no significant spectral difference between the moonlets and the primary. 
 
 \subsection{The (93) Minerva system}
As shown in Figure 4a, the two moons are well separated from the primary in the IFS data and the halo of the primary has been removed effectively using the same method described above. We adopted the effective diameter of the primary of 154$\pm$6 km from \cite{marchis:2013b} and estimated the sizes of the two satellites using the ratios of the integrated fluxes between the satellites and the primary. Assuming the satellites share the same albedos as the primary, the estimated diameter for Aegis is 4.3$\pm$1.5 km and the diameter of Gorgoneion is 3.7$\pm$1.5 km, which are consistent with the previous diameter estimates of 3.6$\pm$1.0 km and 3.2$\pm$0.9 km in \cite{marchis:2013b}. The large differences between the primary spectrum and the satellites' spectra, at wavelengths between 1.3~$\mu$m and 1.5~$\mu$m, are a result of strong atmospheric absorption. We also performed a t-test for the Minerva system. We found no discernible difference between satellite and primary, both for Aegis (P=0.64) and for Gorgoneion (P=0.79).  

\section{Discussion}  
The major challenge in obtaining reliable component-resolved spectroscopy of tight asteroid systems is the bright halo of the primary and the variable sky background. The outer satellites, i.e.\ S/2003 (130) 1 and I Aegis which are orbiting at 13 and 8 primary radii and have detection SNR $>$ 10, yield spectra which are not very sensitive to our image processing methods. In contrast, the spectral slopes of the inner moons (S/2014 (130) 1 and II Gorgoneion) are heavily influenced by the halo removal process, because these two are closer to the primary and are faint in comparison to the background noise. As such, it is necessary to understand the impact of the halo removal methods at different wavelengths. Inserting simulated sources with known PSFs and fluxes into the real dataset and processing the data in the same manner as the science is perhaps the most reliable way to correct photometric biases and estimate uncertainties. 

\renewcommand{\thefootnote}{$\star$} 
Our simulations show large uncertainties in the spectral measurements. Therefore, we are not able to further investigate space weathering effects on the individual components of the triple systems. In terms of the satellites, we detect no absorption band near 1.0 micron and the spectral slopes of these moonlets are flat (within a few percent). Our IFS observations suggest that the components of both the Elektra and the Minerva systems are similar in composition and the four satellites are best classified as C-types. The rotational periods for (130) Elektra and (93) Minerva are 5.225 and 5.982\footnote{Taken from minor planet lightcurve parameters, A.W. Harris and B.D. Warner, http://cfa-www.harvard.edu/iau/lists/LightcurveDat.html.} hours respectively. Therefore, break-up from a high spin rate is unlikely to have led to formation these triple systems. Given the small separations of the satellites from their primaries and the large mass ratios between the moonlets and primaries, an erosive impact is the most likely explanation for both the Elektra and the Minerva triple systems. 

As of now, the eccentricity of the satellite orbits for most multiple asteroid systems have not been estimated because of the difficulty in obtaining sufficient position measurements of the satellites. For the few systems for which we have enough observations to characterize the satellites' orbits, a large range of eccentricities have been observed, see Figure 2b. Our observations of (130) Elektra reveal a unique case: both moonlets are on eccentric orbits. We note that asteroid Elektra is elongated with $a/b \sim$1.5. However, it is known that the elongation of the primary has no long-term effect on the eccentricity, the semi-major axis or inclination \citep{murray:2000}. More detailed dynamical simulations of the orbital evolution of the asteroid triple systems are needed to understand the difference in eccentricity between the inner and the outer moon. 

In summary, our SPHERE observations yield the following findings: 1) a new moonlet of (130) Elektra was detected and the estimated diameter of this moon is $\sim$2km, 2) both the inner and the outer moonlet of (130) Elektra are on orbits with non-negligible eccentricities, 3) the reflectance spectra of the satellites of the (130) Elektra and the (93) Minvera systems are similar to those of their primaries, suggesting that both triple systems were created by disruptive impacts.

\begin{acknowledgements}
We thank Julien Milli and Arthur Vigan for helping with the SPHERE data reduction. Special thank to David Jewitt for commenting on the paper. FM was supported in part by NASA cooperative agreements NNX14AJ80G.
\end{acknowledgements}



\begin{thebibliography}{20}
\expandafter\ifx\csname natexlab\endcsname\relax\def\natexlab#1{#1}\fi

\bibitem[{Beauvalet \& Marchis(2014)}]{beauvalet:2014}
Beauvalet, L., \& Marchis, F. 2014, Icarus, 241, 13

\bibitem[{Beauvalet {et~al.}(2013)Beauvalet, Robert, Lainey, Arlot, \&
  Colas}]{beauvalet:2013}
Beauvalet, L., Robert, V., Lainey, V., Arlot, J.~E., \& Colas, F. 2013,
  A\&A, 553, A14

\bibitem[Berthier et al.(2014)]{berthier:2014} Berthier, J., Vachier, 
F., Marchis, F., {\v D}urech, J., \& Carry, B.\ 2014, Icarus, 239, 118 

\bibitem[Beuzit et al.(2008)]{beuzit:2008} Beuzit, J.-L., Feldt, 
M., Dohlen, K., et al.\ 2008, \procspie, 7014, 701418 

\bibitem[Birlan et al.(2014)]{birlan:2014} Birlan, M., Nedelcu, 
D.~A., Popescu, M., et al.\ 2014, \mnras, 437, 176 

\bibitem[Bus 
\& Binzel(2002)]{bus:2002} Bus, S.~J., \& Binzel, R.~P.\ 2002, Icarus, 158, 146 

\bibitem[Claudi et al.(2008)]{claudi:2008} Claudi, R.~U., Turatto, 
M., Gratton, R.~G., et al.\ 2008, \procspie, 7014, 70143E 

\bibitem[{Cruikshank \& Brown(1987)}]{cruikshank:1987}
Cruikshank, D.~P., \& Brown, R.~H. 1987, Science, 238, 183

\bibitem[{DeMeo {et~al.}(2009)DeMeo, Binzel, Slivan, \& Bus}]{deMeo:2009}
DeMeo, F.~E., Binzel, R.~P., Slivan, S.~M., \& Bus, S.~J. 2009, Icarus, 202,
  160

\bibitem[{DeMeo {et~al.}(2011)DeMeo, Carry, Marchis, Birlan, Binzel, Bus,
  Descamps, Nedelcu, Busch, \& Bouy}]{deMeo:2011}
DeMeo, F.~E., {et~al.} 2011, Icarus, 212, 677

\bibitem[Descamps et al.(2011)]{descamps:2011} Descamps, P., Marchis, 
F., Berthier, J., et al.\ 2011, Icarus, 211, 1022 

\bibitem[Dohlen et al.(2008)]{dohlen:2008} Dohlen, K., Langlois, 
M., Saisse, M., et al.\ 2008, \procspie, 7014, 70143L 

\bibitem[Fang et al.(2012)]{fang:2012} Fang, J., Margot, J.-L., 
\& Rojo, P.\ 2012, \aj, 144, 70 

\bibitem[{Funato {et~al.}(2004)Funato, Makino, Hut, Kokubo, \&
  Kinoshita}]{funato:2004}
Funato, Y., Makino, J., Hut, P., Kokubo, E., \& Kinoshita, D. 2004, Nature,
  427, 518

\bibitem[{Laver {et~al.}(2009)Laver, de~Pater, Marchis, {\'A}d{\'a}mkovics, \&
  Wong}]{laver:2009}
Laver, C., de~Pater, I., Marchis, F., {\'A}d{\'a}mkovics, M., \& Wong, M.~H.
  2009, Icarus, 204, 574

\bibitem[Lantz et al.(2013)]{lantz:2013} Lantz, C., Clark, B.~E., Barucci, M.~A., \& Lauretta, D.~S.\ 2013, \aap, 554, A138 

\bibitem[{Lazzaro {et~al.}(2004)Lazzaro, Angeli, Carvano, Motha-Diniz, Duffard,
  \& Florczak}]{lazzaro:2004}
Lazzaro, D., Angeli, C.~A., Carvano, J.~M., Motha-Diniz, T., Duffard, R., \&
  Florczak, M. 2004, Icarus, 172, 179

\bibitem[Liu(2004)]{liu:2004} Liu, M.~C.\ 2004, Science, 305, 
1442 

\bibitem[Margot et al.(2015)]{margot:2015} Margot, J.-L., Pravec, 
P., Taylor, P., Carry, B., \& Jacobson, S.\ 2015, ``Asteroid Systems: Binaries, Triples, and Pairs",
 chapter in the book ASTEROIDS IV (in press).

\bibitem[{Marchis {et~al.}(2013{\natexlab{a}})Marchis, Ruffio, Vachier, \&
  Berthier}]{marchis:2013a}
Marchis, F., Ruffio, J., Vachier, F., \& Berthier, J. 2013{\natexlab{a}}, AGU
  Fall Meeting Abstracts, 23, 1760

\bibitem[{Marchis {et~al.}(2013{\natexlab{b}})Marchis, Vachier, {\v D}urech,
  Enriquez, Harris, Dalba, Berthier, Emery, Bouy, Melbourne, Stockton,
  Fassnacht, Dupuy, \& Strajnic}]{marchis:2013b}
Marchis, F., {et~al.} 2013{\natexlab{b}}, Icarus, 224, 178

\bibitem[Marchis et al.(2012)]{marchis:2012} Marchis, F., Enriquez, 
J.~E., Emery, J.~P., et al.\ 2012, Icarus, 221, 1130 

\bibitem[Marchis et al.(2011)]{marchis:2011} Marchis, F., Enriquez, 
J.~E., Emery, J.~P., et al.\ 2011, Icarus, 213, 252 

\bibitem[Marchis et al.(2010)]{marchis:2010} Marchis, F., Lainey, 
V., Descamps, P., et al.\ 2010, Icarus, 210, 635 

\bibitem[Marchis et al.(2009)]{marchis:2009}Marchis, F., Macomber, B., Berthier, J., Vachier, F., Emery, J.P. 2009. In: Green, D.W.E.
(Ed.), S/2009 (93)1and S/2009 (93)2. IAU Circ. 9069(1).

\bibitem[Marchis et al.(2008{\natexlab{a}})]{marchis:2008a} Marchis, F., Descamps, 
P., Baek, M., et al.\ 2008a, Icarus, 196, 97 

\bibitem[Marchis et al.(2008{\natexlab{b}})]{marchis:2008b}
Marchis, F., Descamps, P., Berthier, J., Hestroffer, D., Vachier, F., Baek, M.,
  Harris, A.~W., \& Nesvorn{\'y}, D. 2008b, Icarus, 195, 295

\bibitem[Marchis et al.(2006)]{marchis:2006} Marchis, F., 
Kaasalainen, M., Hom, E.~F.~Y., et al.\ 2006, Icarus, 185, 39 

\bibitem[Marois et al.(2006)]{marois:2006} Marois, C., 
Lafreni{\`e}re, D., Doyon, R., Macintosh, B., 
\& Nadeau, D.\ 2006, \apj, 641, 556 

\bibitem[{Mesa {et~al.}(2015)Mesa, Gratton, Zurlo, Vigan, Claudi, Alberi,
  Antichi, Baruffolo, Beuzit, Boccaletti, Bonnefoy, Costille, Desidera, Dohlen,
  Fantinel, Feldt, Fusco, Giro, Henning, Kasper, Langlois, Maire, Martinez,
  Moeller-Nilsson, Mouillet, Moutou, Pavlov, Puget, Salasnich, Sauvage, Sissa,
  Turatto, Udry, Vakili, Waters, \& Wildi}]{mesa:2015}
Mesa, D., {et~al.} 2015, A\&A, 576, A121

\bibitem[Murray 
\& Dermott(2000)]{murray:2000} Murray, C.~D., \& Dermott, S.~F.\ 2000, Solar System Dynamics, by C.D.~Murray and S.F.~Dermott.~ ISBN 0521575974. Cambridge, UK: Cambridge University Press, 2000.

\bibitem[Pavlov et al.(2008)]{pavlov:2008} Pavlov, A., 
M{\"o}ller-Nilsson, O., Feldt, M., et al.\ 2008, \procspie, 7019, 701939 


\bibitem[{Takir \& Emery(2012)}]{takir:2012}
Takir, D., \& Emery, J.~P. 2012, Icarus, 219, 641

\bibitem[Tholen 
\& Barucci(1989)]{tholen:1989} Tholen, D.~J., \& Barucci, M.~A.\ 1989, In: Asteroids II;
Proceedings of the Conference, pp. 298Ð315.

\bibitem[{Usui {et~al.}(2011)Usui, Kuroda, M{\"u}ller, Hasegawa, Ishiguro,
  Ootsubo, Ishihara, Kataza, Takita, Oyabu, Ueno, Matsuhara, \&
  Onaka}]{usui:2011}
Usui, F., {et~al.} 2011, PASJ, 63,
  1117

\bibitem[Vachier et 
al.(2012)]{vachier:2012} Vachier, F., Berthier, J., \& Marchis, F.\ 2012, \aap, 543, A68 

\bibitem[{Vigan {et~al.}(2015)Vigan, Gry, Salter, Mesa, Homeier, Moutou, \&
  Allard}]{vigan:2015}
Vigan, A., Gry, C., Salter, G., Mesa, D., Homeier, D., Moutou, C., \& Allard,
  F. 2015, MNRAS, 454, 129

\bibitem[Vigan et al.(2010)]{vigan:2010} Vigan, A., Moutou, C., 
Langlois, M., et al.\ 2010, \mnras, 407, 71 

\bibitem[{Wahhaj {et~al.}(2013)Wahhaj, Liu, Biller, Nielsen, Close, Hayward,
  Hartung, Chun, Ftaclas, \& Toomey}]{wahhaj:2013}
Wahhaj, Z., {et~al.} 2013, ApJ, 779, 80

\end{thebibliography}

\begin{deluxetable}{lllcccccc}\tablewidth{5.0in}{!hb}
\tabletypesize{\scriptsize}
\hspace{-25 cm}
\tablecaption{Astrometric and Photometric Measurements for the (93) Minerva and (130) Elektra Triple Systems
  \label{obstable}}
\tablecolumns{12} \tablehead{ \colhead{Object}  & \colhead{Date } & \colhead{Time } & \colhead{$\delta$X\tablenotemark{a}}& \colhead{$\delta$Y\tablenotemark{a}}  & \colhead{$\delta$D\tablenotemark{b}}& \colhead{$\Delta$m \tablenotemark{c}}  & \colhead{D$_{S}$\tablenotemark{d}} \\
	\colhead{}&\colhead{(UT)}&\colhead{(UT)}&\colhead{(arcsec)}&\colhead{(arcsec)}&\colhead{(arcsec)}&\colhead{(mag)}&\colhead{(km)}}
\startdata
S/2003 (130) 1 & 2014-Dec-06 & 03:25:22 & -0.351$\pm$0.002 & -0.625$\pm$0.004 & 0.717 & 8.2 & 4.4 \\
S/2003 (130) 1 & 2014-Dec-06 & 03:36:20 & -0.339$\pm$0.002 & -0.624$\pm$0.002 & 0.710 & 8.0 & 4.7 \\
S/2003 (130) 1 & 2014-Dec-06 & 03:47:43 & -0.327$\pm$0.003 & -0.623$\pm$0.004 & 0.703 & 8.0 & 4.8 \\
S/2003 (130) 1 & 2014-Dec-09 & 01:33:19 & 0.317$\pm$0.001 & 0.555$\pm$0.003 & 0.639 & 7.6 & 5.8 \\
S/2003 (130) 1 & 2014-Dec-09 & 01:36:13 & 0.315$\pm$0.001 & 0.557$\pm$0.003 & 0.640 & 7.7 & 5.8 \\
S/2003 (130) 1 & 2014-Dec-09 & 01:39:11 & 0.313$\pm$0.001 & 0.557$\pm$0.003 & 0.639 & 7.7 & 5.7 \\
S/2003 (130) 1 & 2014-Dec-09 & 01:42:05 & 0.309$\pm$0.001 & 0.554$\pm$0.005 & 0.635 & 7.6 & 5.8 \\
S/2003 (130) 1 & 2014-Dec-09 & 01:45:00 & 0.307$\pm$0.001 & 0.557$\pm$0.002 & 0.637 & 7.6 & 5.9 \\
S/2003 (130) 1 & 2014-Dec-09 & 01:47:56 & 0.305$\pm$0.001 & 0.558$\pm$0.001 & 0.636 & 7.6 & 6.0 \\
S/2003 (130) 1 & 2014-Dec-09 & 01:50:50 & 0.302$\pm$0.001 & 0.554$\pm$0.004 & 0.631 & 7.6 & 5.9 \\
S/2003 (130) 1 & 2014-Dec-09 & 01:53:45 & 0.301$\pm$0.001 & 0.558$\pm$0.001 & 0.634 & 7.6 & 6.1 \\
S/2003 (130) 1 & 2014-Dec-09 & 01:56:40 & 0.298$\pm$0.001 & 0.556$\pm$0.003 & 0.631 & 7.6 & 6.0 \\
S/2003 (130) 1 & 2014-Dec-09 & 01:59:36 & 0.299$\pm$0.003 & 0.565$\pm$0.006 & 0.639 & 7.6 & 6.0 \\
S/2003 (130) 1 & 2014-Dec-30 & 01:05:42 & 0.491$\pm$0.002 & 0.456$\pm$0.003 & 0.670 & 7.5 & 6.1 \\
S/2003 (130) 1 & 2014-Dec-30 & 01:29:57 & 0.475$\pm$0.002 & 0.458$\pm$0.004 & 0.660 & 7.6 & 6.1 \\
S/2003 (130) 1 & 2014-Dec-30 & 01:39:00 & 0.467$\pm$0.002 & 0.458$\pm$0.003 & 0.654 & 7.6 & 6.0 \\
S/2003 (130) 1 & 2014-Dec-31 & 03:45:33 & -0.683$\pm$0.002 & 0.120$\pm$0.003 & 0.693 & 7.8 & 5.4 \\
S/2003 (130) 1 & 2014-Dec-31 & 03:54:35 & -0.688$\pm$0.001 & 0.117$\pm$0.005 & 0.698 & 7.7 & 5.6 \\
S/2003 (130) 1 & 2014-Dec-31 & 04:03:35 & -0.691$\pm$0.002 & 0.112$\pm$0.003 & 0.700 & 7.6 & 5.8 \\
 \hline\\
S/2014 (130) 1 & 2014-Dec-06 & 03:25:22 & -0.374$\pm$0.005 & -0.062$\pm$0.006 & 0.379 & 9.99 & 1.7\\
S/2014 (130) 1 & 2014-Dec-06 & 03:36:20 & -0.369$\pm$0.005 & -0.073$\pm$0.005 & 0.376 & 9.95 & 1.8\\
S/2014 (130) 1 & 2014-Dec-06 & 03:47:43 & -0.359$\pm$0.007 & -0.080$\pm$0.008 & 0.367 & 10.71 & 1.3\\
S/2014 (130) 1 & 2014-Dec-09 & 01:33:19 & 0.361$\pm$0.005 & -0.002$\pm$0.005 & 0.361 & 9.74 & 2.2\\
S/2014 (130) 1 & 2014-Dec-09 & 01:36:13 & 0.361$\pm$0.004 & 0.001$\pm$0.005 & 0.361 & 9.74 & 2.2\\
S/2014 (130) 1 & 2014-Dec-09 & 01:39:11 & 0.363$\pm$0.003 & 0.004$\pm$0.004 & 0.363 & 9.77 & 2.2\\
S/2014 (130) 1 & 2014-Dec-09 & 01:42:05 & 0.362$\pm$0.003 & 0.003$\pm$0.006 & 0.362 & 9.66 & 2.3\\
S/2014 (130) 1 & 2014-Dec-09 & 01:45:00 & 0.365$\pm$0.003 & 0.010$\pm$0.003 & 0.365 & 9.66 & 2.3\\
S/2014 (130) 1 & 2014-Dec-09 & 01:47:56 & 0.366$\pm$0.003 & 0.011$\pm$0.003 & 0.366 & 9.76 & 2.2\\
S/2014 (130) 1 & 2014-Dec-09 & 01:50:50 & 0.365$\pm$0.003 & 0.013$\pm$0.004 & 0.365 & 9.75 & 2.2\\
S/2014 (130) 1 & 2014-Dec-09 & 01:53:45 & 0.367$\pm$0.003 & 0.018$\pm$0.003 & 0.368 & 9.77 & 2.2\\
S/2014 (130) 1 & 2014-Dec-09 & 01:56:40 & 0.370$\pm$0.003 & 0.020$\pm$0.004 & 0.371 & 9.78 & 2.2\\
S/2014 (130) 1 & 2014-Dec-09 & 01:59:36 & 0.375$\pm$0.004 & 0.032$\pm$0.007 & 0.377 & 9.90 & 2.1\\
S/2014 (130) 1 & 2014-Dec-30 & 01:05:42 & -0.322$\pm$0.004 & -0.103$\pm$0.005 & 0.338 & 9.70 & 2.3\\
S/2014 (130) 1 & 2014-Dec-30 & 01:29:57 & -0.320$\pm$0.006 & -0.130$\pm$0.007 & 0.337 & 9.59 & 2.4\\
S/2014 (130) 1 & 2014-Dec-30 & 01:39:00 & -0.307$\pm$0.003 & -0.141$\pm$0.004 & 0.338 & 9.25 & 2.8\\
S/2014 (130) 1 & 2014-Dec-31 & 03:45:33 & -0.315$\pm$0.011 & -0.017$\pm$0.011 & 0.316 & 10.46 & 1.6\\
S/2014 (130) 1 & 2014-Dec-31 & 03:54:35 & -0.316$\pm$0.009 & -0.018$\pm$0.010 & 0.317 & 10.76 & 1.4\\
S/2014 (130) 1 & 2014-Dec-31 & 04:03:35 & -0.316$\pm$0.004 & -0.034$\pm$0.005 & 0.318 & 10.17 & 1.9\\
 \hline\\
 Aegis                 & 2014-Dec-08  & 01:48:17& -0.323$\pm$0.001&0.085$\pm$0.002  & 0.333 & 7.25 & 4.3 \\
Gorgoneion      & 2014-Dec-08  & 01:48:17& -0.037$\pm$0.002 &-0.231$\pm$0.002 & 0.234 & 8.65 & 3.7 \\
\enddata
\tablenotetext{a}{The relative cartesian coordinates ($\delta$X and $\delta$Y) of the satellite with respect to the optical center of the primary.}
\tablenotetext{b}{Angular separation of the satellite from the primary.}
\tablenotetext{c}{Integrated flux ratio between the primary and the satellite.}
\tablenotetext{d}{Estimated diameter of the satellite using the integrated flux ratio of the primary and the satellite, assuming the same albedo.}

\end{deluxetable}

\begin{figure}[hb!]
\begin{center}
\vspace{1.5 cm}
\hspace{-0.5 cm}\includegraphics[width=7.2in,angle=0]{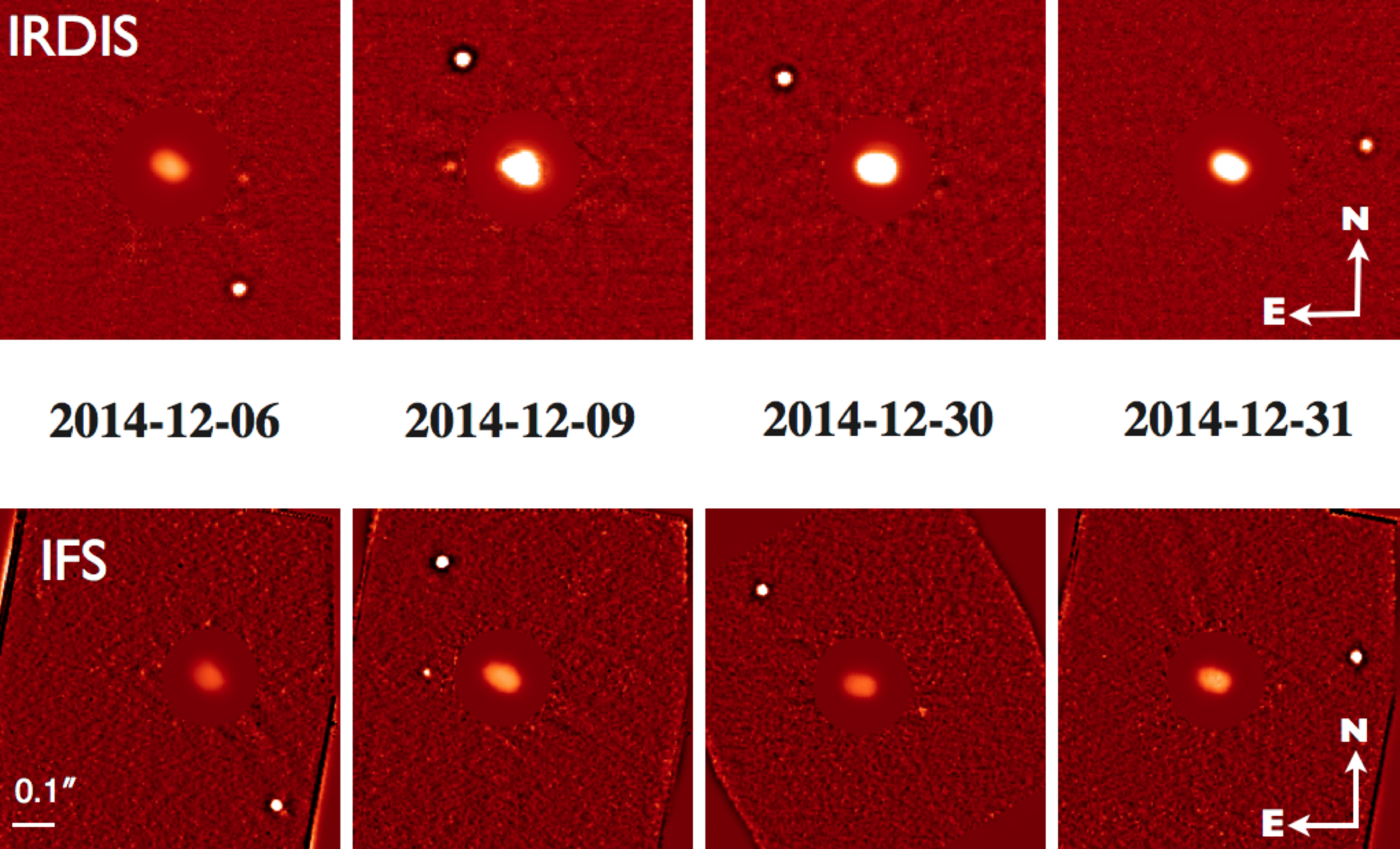}
\caption{Reduced IRDIS and IFS images of (130) Elektra, the triple asteroid system. The IFS images are the median combination of 31 of the 39 spectral channels. The channels severely affected by the atmospheric absorption were not included. The IFS images are rotated by 100.7 degree clockwise to align with the corresponding IRIDS data with PA=0. The pixels intensities within 0.25$''$ of the primary have been divided by 1500 to make the faint satellites more easily viewable. A new satellite S/2014 (130) S1 was detected in both the IRDIS and IFS data sets in most epochs. This inner moon detection was weaker in the December 31 data and was only identified in the IFS data.}
\end{center}
\label{fig1}
\end{figure}

\clearpage
\begin{figure}[!ht]
\begin{center}
\vspace{2 cm}
\includegraphics[width=3.5in,angle=0]{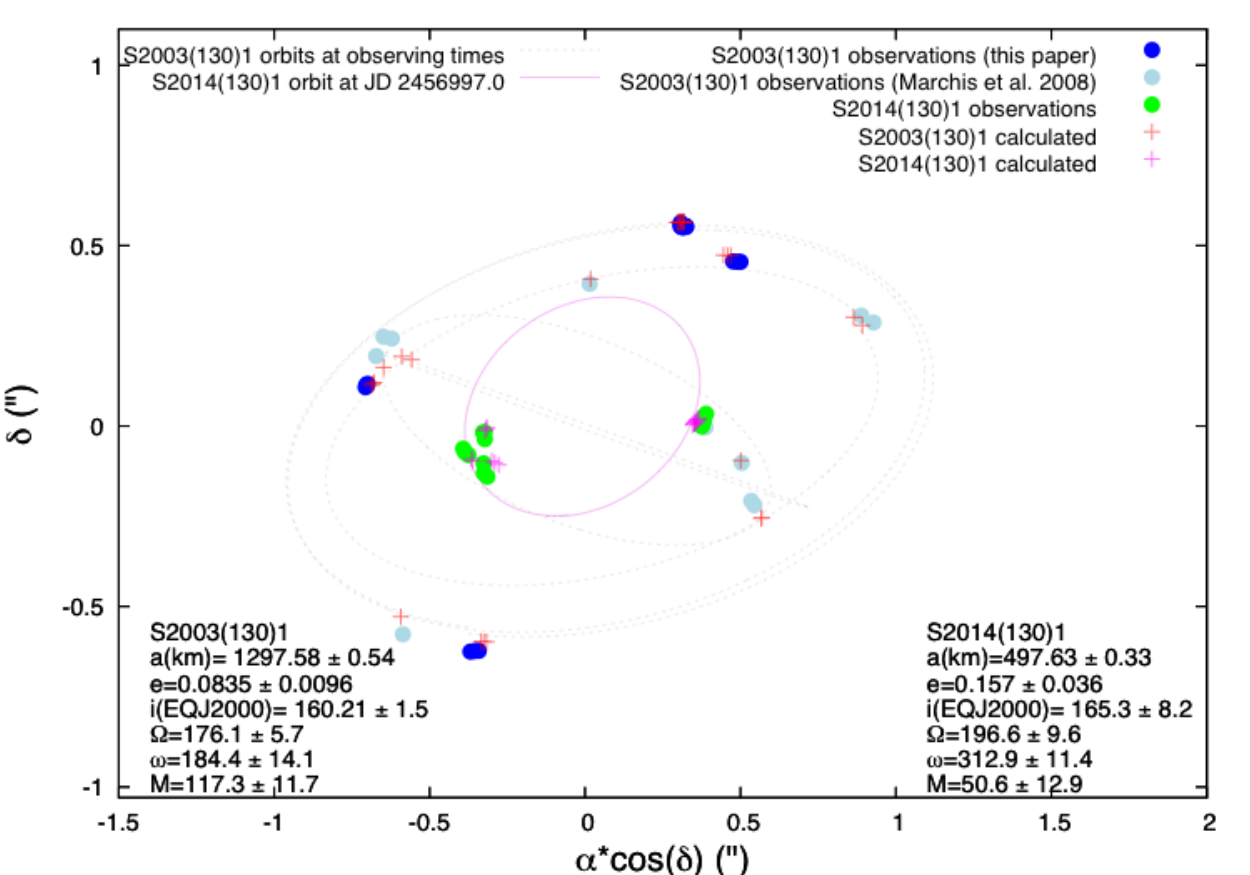}\includegraphics[width=3.55in,angle=0]{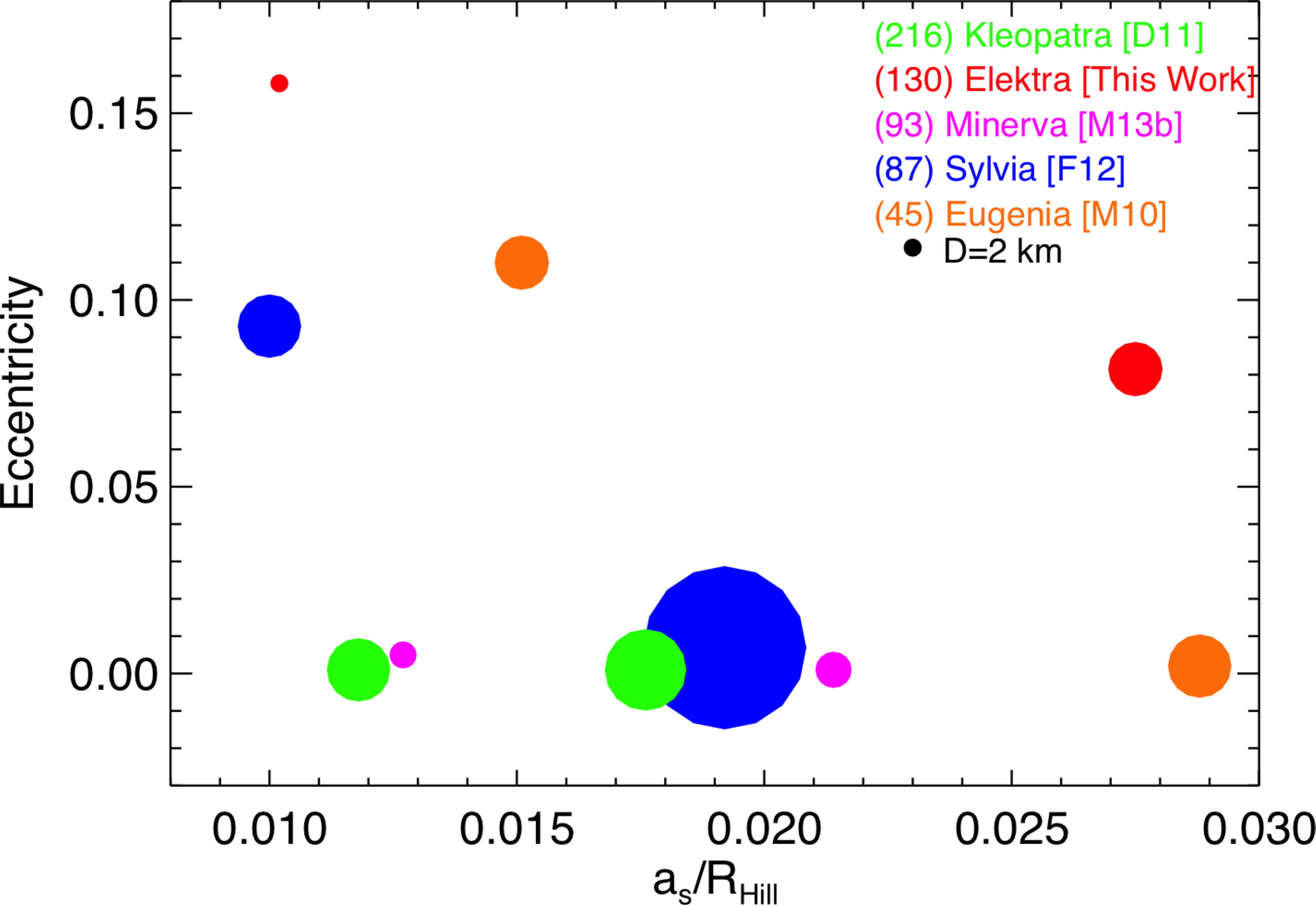}
\caption{Left: the grey dashed lines show the best-fit orbit for S/2003 S1, as it evolved over many epochs, and the pink line is the best-fit orbit for S/2014 S1. 
The blue and green solid circles are the astrometric measurements of the two moonlets. The red and pink pluses are the simulated positions using the dynamical model ODIN. The uncertainties of the orbital elements are 3-$\sigma$ errors derived using the random hold-out method. Right: the eccentricities of the satellites that belong to the 5 known triple systems; $a_s$ is the semi-major axis of the satellite's orbit and $R_{Hill}$ is the Hill radius of the primary. References: \cite{marchis:2010}[M10], \cite{descamps:2011}[D11], \cite{fang:2012}[F12], \cite{marchis:2013b}[M13b].}
\label{fig2}
\end{center}
\end{figure}

\begin{figure}[h]
\vspace{2 cm}
\includegraphics[width=3.5in,angle=0]{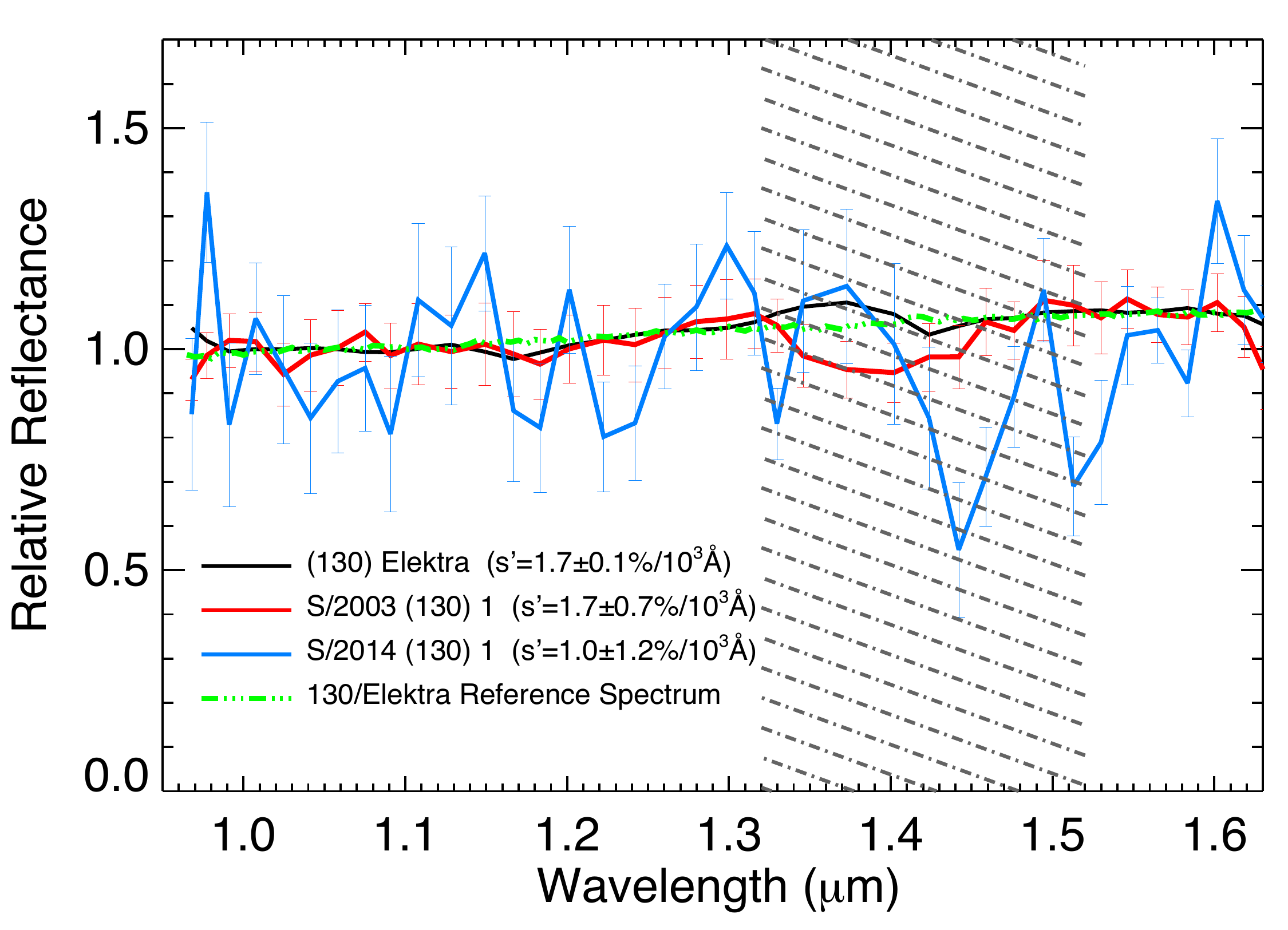}\includegraphics[width=3.5in,angle=0]{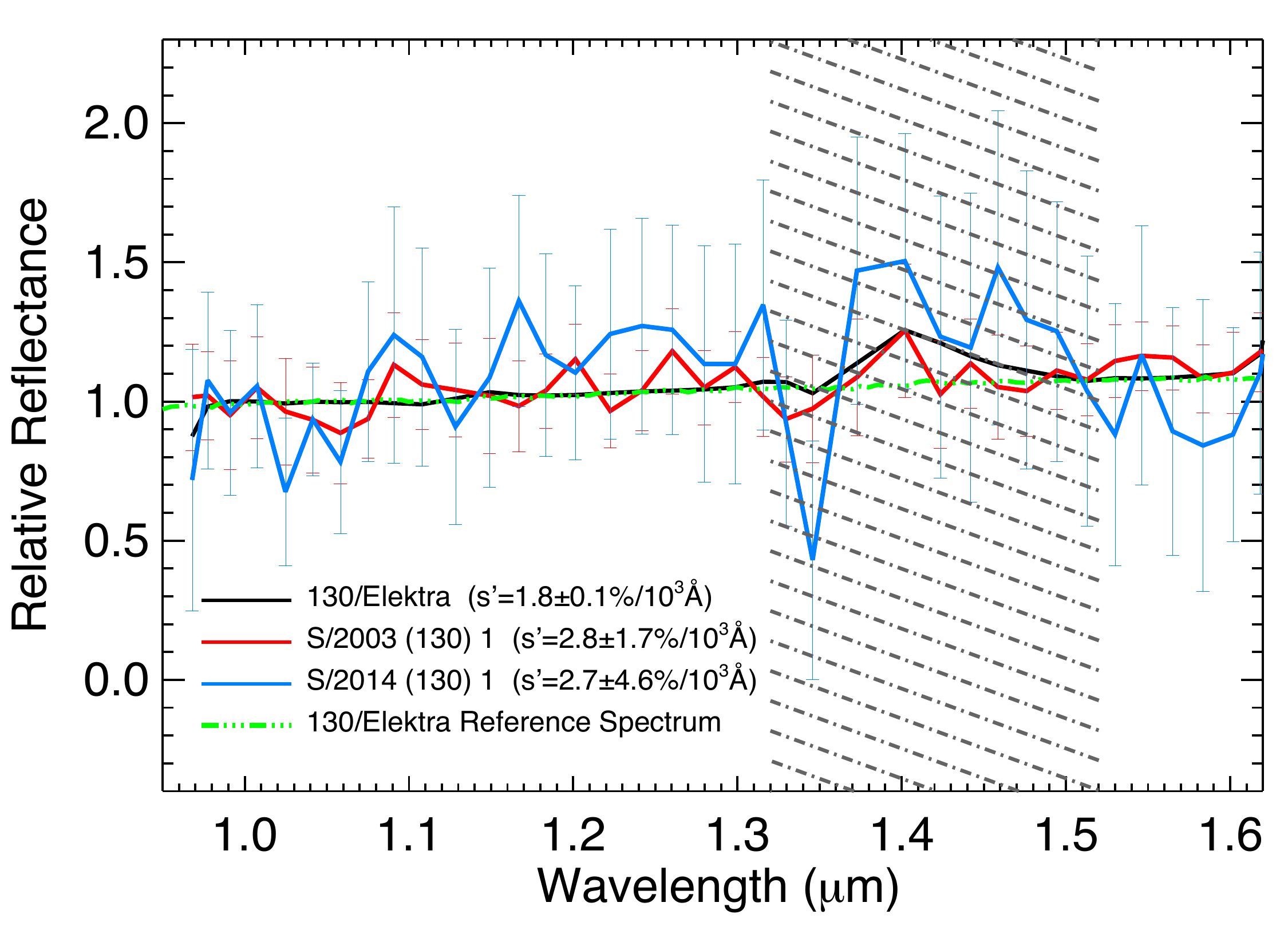}\\
\caption{The relative reflectance spectra of the (130) Elektra system, normalized at 1.0 $\mu$m. Left:  Spectra of the two satellites and the primary obtained on UT December 09, using PSF fitting. Right: Spectra of the two satellites and the primary obtained on UT December 30. The spectrum of the inner moon, obtained via aperture photometry. The spectra of the primary and the outer moon were obtained via PSF fitting. For comparison, we also plot the near infrared spectrum of the primary obtained with the 3.0 m IRTF telescope by \cite{takir:2012}. The wavelength range affected by telluric absorption is shaded in the plot. }
\label{fig3}
\end{figure}

\begin{figure}[h]
\begin{center}
\vspace{1.8 cm}
\includegraphics[width=3.0in,angle=0]{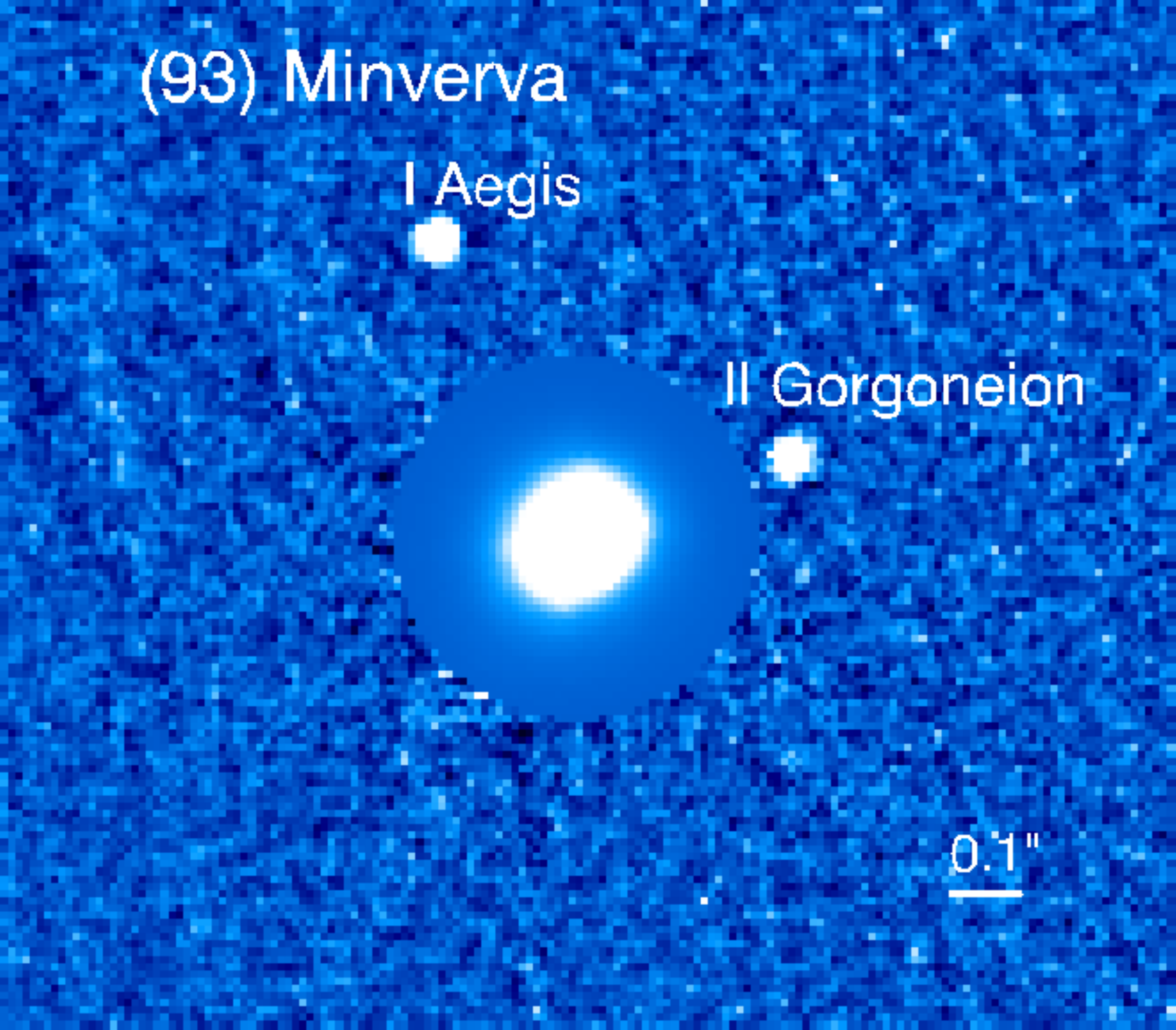} \includegraphics[width=3.8in,angle=0]{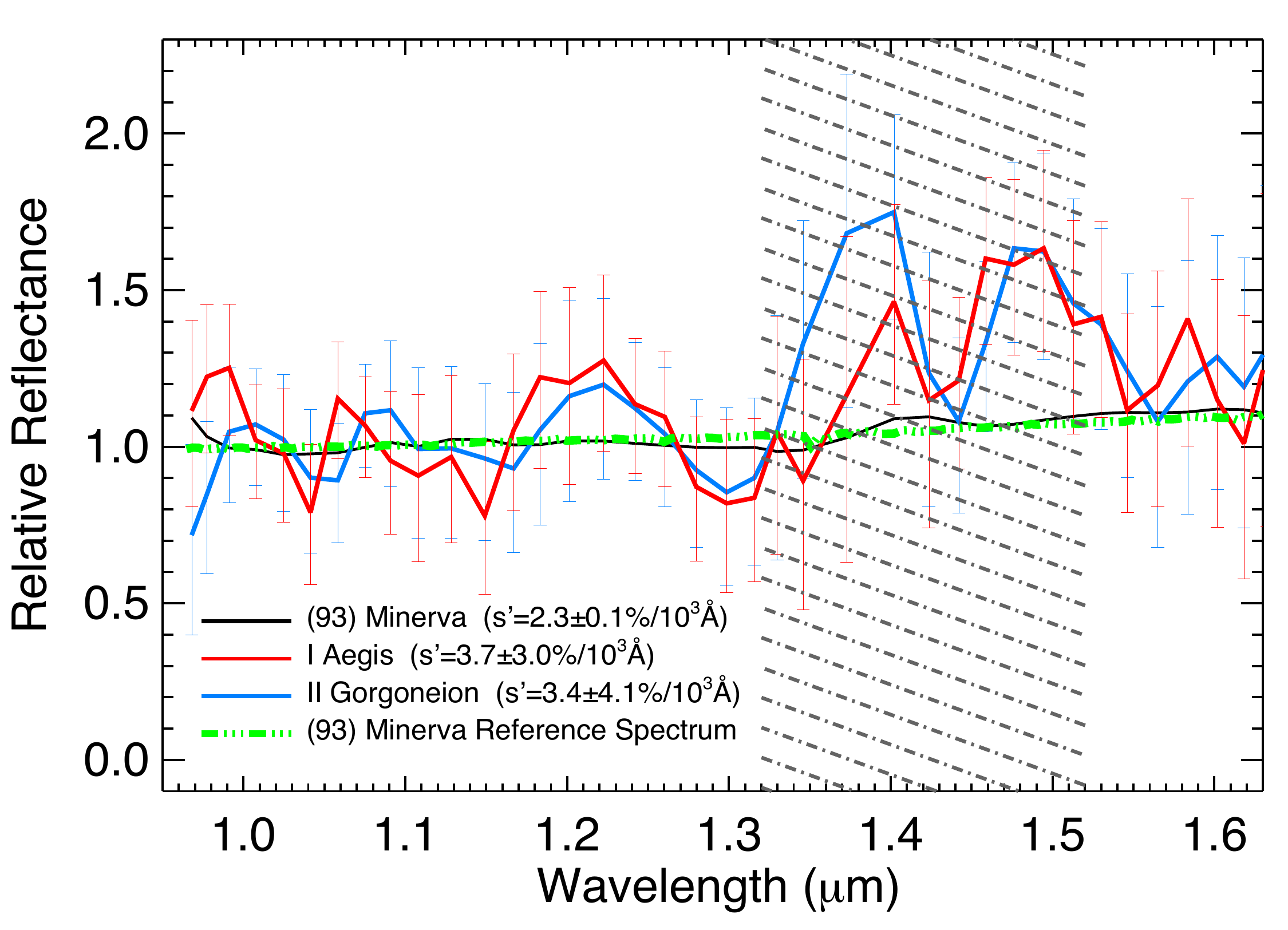}
\caption{Left: Reduced IFS image of the Minerva system, which is  the median combination of 31 of the 39 spectral channels. The pixel intensities within 0.25$''$ of the primary have been divided by 300 to show the two satellites clearly. Right:  The relative reflectance spectra of Aegis and Gorgoneion were obtained via aperture photometry, normalized at 1.0~$\mu$m. The reflectance spectra of the individual components of (93) Minerva are presented in Figure 4b. A comparison IRTF spectrum of the primary obtained by \cite{marchis:2013b} is shown in green.}
\label{fig4}
\end{center}
\end{figure}

\clearpage

\end{document}